\documentclass[aps,prb,superscriptaddress,twocolumn,showpacs,amsmath,floatfix,citeautoscript]{revtex4}

\usepackage{graphicx}
\usepackage{float}
\usepackage{dcolumn}
\usepackage{color}
\usepackage{latexsym,bm}
\usepackage[normalem]{ulem}
\usepackage{multirow}

\begin{document}

\hyphenpenalty=5000

\tolerance=1000

\title{Reply to comments on ``Gapless spin liquid ground state of spin-1/2 $J_1$-$J_2$  Heisenberg  model on square lattices''}

\author{Wen-Yuan Liu}
\altaffiliation{Current address: Department of Physics, The Chinese University of Hong Kong, Shatin, New Territories, Hong Kong, China}
\affiliation{Key Laboratory of Quantum Information, University of Science and
  Technology of China, Hefei, Anhui, 230026, People's Republic of China}
\affiliation{Synergetic Innovation Center of Quantum Information and Quantum
  Physics, University of Science and Technology of China, Hefei, 230026, China}

\author{Shaojun Dong}
\affiliation{Key Laboratory of Quantum Information, University of Science and
  Technology of China, Hefei, Anhui, 230026, People's Republic of China}
\affiliation{Synergetic Innovation Center of Quantum Information and Quantum
  Physics, University of Science and Technology of China, Hefei, 230026, China}

\author{Chao Wang}
\affiliation{Key Laboratory of Quantum Information, University of Science and
  Technology of China, Hefei, Anhui, 230026, People's Republic of China}
\affiliation{Synergetic Innovation Center of Quantum Information and Quantum
  Physics, University of Science and Technology of China, Hefei, 230026, China}

\author{Yongjian Han}
\email{smhan@ustc.edu.cn}
\affiliation{Key Laboratory of Quantum Information, University of Science and
  Technology of China, Hefei, Anhui, 230026,  People's Republic of China}
\affiliation{Synergetic Innovation Center of Quantum Information and Quantum
  Physics, University of Science and Technology of China, Hefei, 230026, China}

\author{Hong An}
\affiliation{School of Computer Science and Technology,University of Science and Technology of China, Hefei, 230026,
  China}

\author{Guang-Can Guo}
\affiliation{Key Laboratory of Quantum Information, University of Science and
  Technology of China, Hefei, Anhui, 230026,  People's Republic of China}
\affiliation{Synergetic Innovation Center of Quantum Information and Quantum
  Physics, University of Science and Technology of China, Hefei, 230026,
  China}

\author{Lixin He}
\email{helx@ustc.edu.cn}
\affiliation{Key Laboratory of Quantum Information, University of Science and
  Technology of China, Hefei, Anhui, 230026,  People's Republic of China}
\affiliation{Synergetic Innovation Center of Quantum Information and Quantum
  Physics, University of Science and Technology of China, Hefei, 230026, China}
\date{\today }
\maketitle

\date{\today}

In a recent comments\cite{zhao2019}, Zhao et al. argue that the definition of dimer orders used in our paper (Ref.~\onlinecite{liu2018}) may not rule out  valence bond solid (VBS) orders of $J_1$-$J_2$ model on the open boundary conditions (OBC). In this reply, we show that their argument does not apply to our case.

The definition of dimer order parameter in our paper~\cite{liu2018} is:
\begin{equation}
m^2_{d\alpha}=\frac{1}{N_b^2}\sum_{{\bf i},{\bf j}}(\langle B^{\alpha}_{\bf i}B^{\alpha}_{\bf j}\rangle-\langle B^{\alpha}_{\bf i}\rangle \langle B^{\alpha}_{\bf j} \rangle) {\rm e}^{i{\bf k \cdot (i-j)}},
\label{eq:OriginDef1}
\end{equation}
where $B^{\alpha}_{\bf i}=\bf{S_i}\cdot \bf{S_{i+e_{\alpha}}}$ is the bond operator defined on a pair of nearest neighbour sites ${\bf i}$ and ${\bf i+e_{\alpha}}$ along the $\alpha$ direction with $\alpha=x$ or $y$. Horizontal dimer values $m^2_{dx}$ and vertical ones $m^2_{dy}$ are obtained with ${\bf k}_x=(\pi,0)$ and ${\bf k}_y=(0,\pi)$, respectively. The Eq.~\ref{eq:OriginDef1} can be also expressed as,
\begin{equation}
m^2_{d\alpha}=\langle D^{2}_{\alpha} \rangle-\langle  D_{\alpha} \rangle^2.
\label{eq:OriginDef2}
\end{equation}
where $D_{\alpha}=\frac{1}{N_b}\sum_{\bf i}(-1)^{i_\alpha}B^{\alpha}_{\bf i}$.
It has been analytically proven that the dimer order parameter is nonzero in the thermodynamic limit for typical VBS states\cite{mambrini2006}, and it has been used to detect VBS orders on periodic boundary conditions (PBC)\cite{mambrini2006}, as well as  cylindrical geometryies\cite{jiang2012,gong2014} in numerical simulations.

  \begin{figure}
 \centering
 \includegraphics[width=3.2in]{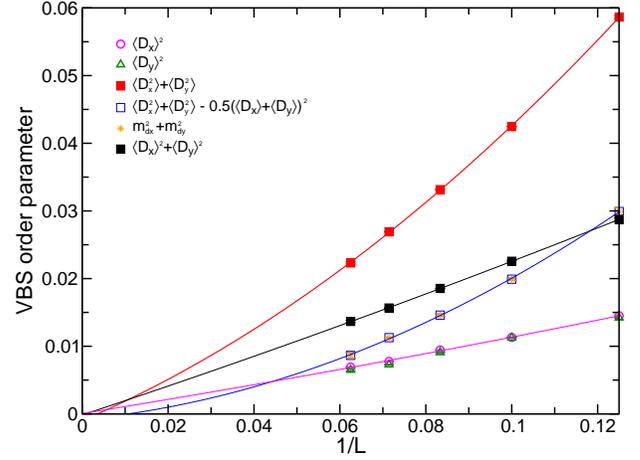}
 \caption{Inverse system size dependence of different definitions of
the %columnar
 VBS order parameter at $J_2/J_1$=0.55. }
 \label{fig:VBSorder}
 \end{figure}

In the comments of Ref.\onlinecite{zhao2019}, Zhao et al. argue that $m^2_{d\alpha}$ can not be used to rule out VBS orders.
They demonstrated their argument on the so-called $J-Q_3$ (with $J=0$) model on open square lattices, whose ground state is a strong VBS state. They propose that one should use other VBS order parameters, such as $\langle D^2_{\alpha}\rangle$, $\langle D_\alpha\rangle^2$, or a symmetrized form  $\langle D_{x}^2\rangle+ \langle D_{y}^2\rangle-{1\over 2}(\langle D_{x}\rangle+\langle D_{y}\rangle)^2$. However, their arguments have a prerequisite, i.e. the ground state has symmetry breaking. In our paper~\cite{liu2018}, we have checked very carefully that there is no symmetry breaking in the ground state. In fact, we have pointed out in our paper~\cite{liu2018}  that ``We find that $m^2_{dx}$ and $m^2_{dy}$ are almost the same within numerical precision at each lattice size, reflecting the isotropy of horizontal and vertical directions, which is expected for the true ground states and exclude the CVB phases.''

To directly address Zhao et al.'s concerns, we plot the VBS order parameters of various definitions including $m^2_{dx}$+$m^2_{dy}$, $\langle D^2_{x}\rangle$+$\langle  D^2_{y}\rangle$,
and $\langle D_{x}^2\rangle+\langle D_{y}^2\rangle-{1\over 2}(\langle D_{x}\rangle+\langle D_{y}\rangle)^2$ etc. at $J_2/J_1$=0.55
up to a $16 \times 16$ lattice~\cite{liu2018} in Fig.~\ref{fig:VBSorder}. As one can see that all the different definitions of VBS order parameters approach zero in the thermodynamic limit by a second order polynomial fitting.
Especially, $\langle D_{x}\rangle^2$ and $\langle D_{y}\rangle^2$　~are almost identical, and as a consequence
the symmetrized order parameters $\langle D_{x}^2\rangle+
\langle D_{y}^2\rangle-{1\over 2}(\langle D_{x}\rangle+\langle D_{y}\rangle)^2$
are identical to $m^2_{dx}$+$m^2_{dy}$. These are strong evidences to rule out the VBS states based on our current calculations.

 In addition, our conclusion that the intermediate nonmagnetic phase of $J_1$-$J_2$ model is a gapless spin liquid state is based on a series of consistent evidences. 
Besides the dimer order parameters, we have also showed that the spin correlation functions have power law decays, suggesting the states have zero ${\bf S}$=1 gap. These results are contradict to those of the VBS states in which the gap is non-zero, and exponential decays of the spin correlation functions are expected.  Importantly, the behaviours at $J_2/J_1$=0.55 are almost the same as those of $J_2/J_1$=0.5, suggesting that no phase transition happens in this region.

We note some progress on tensor network algorithms was reported in a very recent paper\cite{liu2019}. It is expected that the $J_1$-$J_2$ model up to 24$\times$24 will be further investigated in the near future, to reexamine all different scenarios among the calculations by tensor network methods\cite{wang2016,reza2017,liu2018}, the density matrix renormailization group (DMRG) method\cite{jiang2012,gong2014,wang2018}, as well as variational quantum Monte Carlo methods\cite{hu2013}.

%\bibliography{replyTocomments.bib}

\end{document}